\documentclass[submission,copyright,creativecommons]{eptcs}
\providecommand{\event}{GASCom 2024} 

\usepackage[T1]{fontenc}
\usepackage{hyperref}
\usepackage{amsmath}
\usepackage{amssymb}
\usepackage{amsthm}
\usepackage{graphicx}
\usepackage[rgb]{xcolor}
\usepackage{pifont}
\usepackage{stmaryrd}
\usepackage{cite}

\usepackage{bm}

\definecolor{myred}{rgb}{0.9,0,0}
\definecolor{mygre}{rgb}{0,0.5,0}
\definecolor{myblu}{rgb}{0.1,0.2,0.8}
\definecolor{mywhite}{rgb}{1,1,1}
\definecolor{myblack}{rgb}{0,0,0}
\definecolor{myoran}{rgb}{1,.4,.2}
\definecolor{mygray}{rgb}{.3,.3,.3}
\definecolor{mygreDark}{rgb}{0,0.33,0}

\newcommand{\framid}{\mathrm{Id}_{\rm f}}
\newcommand{\aut}{\mathrm{Aut}}

\newcommand{\eee}{\mathbb{E}}

\renewcommand{\aa}{\alpha}

\newcommand{\deenne}[2]{\frac{\partial^#2}{\partial #1 ^#2}}

\newcommand{\be}{\begin{equation}}
\newcommand{\ee}{\end{equation}}
\newcommand{\ef}[1]{\, #1}

\newcommand{\YYno}[1]{$\blacksquare$}
\newcommand{\YYblu}[1]{\textcolor{myblu}{$\blacksquare$}}

\newcommand{\dx}[1] {\mathrm{d}{#1}}

\def\psibar{{\bar{\psi}}}

\let\epsilon\varepsilon

\newcommand{\cG}{\mathcal{G}}

\newcommand{\cO}{\mathcal{O}}

\newcommand{\cD}{\mathcal{D}}

\newcommand{\cL}{\mathcal{L}}

\newcommand{\cB}{\mathcal{B}}

\newcommand{\bZ}{\mathbb{Z}}

\newcommand{\ppp}{\mathbb{P}}

\newcommand{\lam}{\lambda}

\title{Natural Measures on Polyominoes\\ Induced by the Abelian Sandpile Model}
\author{Andrea Sportiello
\institute{CNRS, and LIPN, 
Universit\'e Sorbonne Paris Nord\thanks{The work of A.~Sportiello is
  supported by the French ANR projects
    DIMERS ({\scriptsize ANR-18-CE40-0033}) and 
    COMBINE ({\scriptsize ANR-19-CE48-0011}).}\\
Villetaneuse, France}
\email{andrea.sportiello@univ-paris13.fr}}
\def\titlerunning{Natural Measures on Polyominoes Induced by the Abelian Sandpile Model}
\def\authorrunning{A. Sportiello}
\begin{document}
\maketitle

\begin{abstract}
We introduce a natural Boltzmann measure over polyominoes induced by
boundary avalanches in the Abelian Sandpile Model. Through the study
of a suitable associated process, we give an argument suggesting that
the probability distribution of the avalnche sizes has a power-law
decay with exponent $\frac{3}{2}$, in contrast with the present
understanding of bulk avalanches in the model (which has some exponent
between $1$ and $\frac{5}{4}$), and to the ordinary generating
function of polyominoes (which is conjectured to have a logarithmic
singularity, i.e.\ exponent $1$). We provide some numerical evidence
for our claims, and evaluate some other statistical observables on our
process, most notably the density of triple points.
\end{abstract}

\section{Non-uniform measures on polyominoes from Statistical
  Mechanics}
\label{sec.intro}

\noindent
Given a periodic tiling of the plane, a (general) \emph{polyomino} is
a finite connected geometric structure formed by joining one or more
cells of the tiling edge to edge. The name polyomino is typically
associated to the square grid, while for the triangular and hexagonal
grids the names polyiamonds and polyhexes (respectively) are sometimes
used \cite{unlibro1,trina1,trina2}.

The history in the study of polyominoes started within recreational
mathematics more than one century ago
\cite{guttmannLNP,polibook,gardner}.  In a modern vision, they form a
challenging problem in Combinatorics and Statistical Mechanics (see
e.g.\ \cite[sec.\;10.8]{harpalm} or \cite{altrolibro}),
somewhat in analogy with the study of Self-Avoiding Walks:
despite allowing for an elementary and natural
definition, very little is known rigorously from a mathematical
perspective, although mathematicians and physicists have provided
numerous conjectures that are believed to be true and are strongly
supported by numerical simulations. A reason for this difficulty is
that, within the field of exactly-solvable models in Statistical
Mechanics, we know more about locally-homogeneous random systems, than
about finite compact random structures embedded in Euclidean space.

Given a polyomino $P$, define $n(P)$, the \emph{size} of $P$, as the
number of faces contained in $P$.  The exhaustive generation of
polyominoes, or their enumeration, at any finite size $n$, is a finite
problem, and the problem of decreasing the computational cost of the
associated algorithms has been studied by several authors
\cite{hexarev1,hexarev2,whitt,unalgohex2,jensen2001,jensenG2001,jensenLNP}. 
That is, calling $A_n^{\cL}$ the number of polyominoes of size $n$ on
a given lattice $\cL$, the problem of determining the first $N$ values
$\{A_1^{\cL}, \ldots, A_N^{\cL}\}$, with the smallest possible
asymptotic growth of the complexity as a function of $N$ (and the
largest possible value of $N$ given the present technology), is an
interesting problem in the theory of algorithms,
and also a topic appropriated for GASCom, but it is not
our subject today.

Determining the asymptotic of $A_n^{\cL}$ as a function of $n$ is a
very interesting subject. Pad\'e approximants can be used on the list
of the first few values, so that the previous question is important
also for this goal, but also other insights can give access to this
information, mostly coming from Statistical Physics.  It is believed
that
$A_n^{\cL} \sim c_{\cL}\; \lam_{\cL}^n\,/n$, 
\cite{jensen2001}, where the overall constant $c_{\cL}$ and the 
\emph{growth rate} $\lam_{\cL}$ are expected to depend on the lattice
(for the square lattice it is known that $4.00253 \leq \lam_{\square}
\leq 4.5252$ and the best estimates are $c_{\square} \simeq 0.3169$
and $\lam_{\square} \simeq 4.0626$ \cite{jensenG2001}), while,
crucially, the exponent $-1$ of the algebraic correction is an exact
rational, and it is expected to be \emph{universal} (in the sense of
universality for Critical Phenomena \cite{jzj}), and is a
\emph{critical exponent}, i.e., among its various properties of
robustness, it shall be the same for all two-dimensional lattices.

Finally, it is of interest to determine the asymptotics for large $n$
of statistical observables of large random polyominoes, taken with the
uniform measure. Some examples of interesting observables are the
perimeter, that is, the number of edges on the boundary, and the
gyration radius, that is, the radius of the smallest disk that
contains the polyomino. The average of both these quantities is
expected to scale algebraically with $n$, again with some
\emph{critical exponents} expected to be the same for all lattices.


An interesting subclass of polyominoes consists of
\emph{simply-connected} polyominoes, that is, polyominoes such that
the boundary consists of a single cycle (or, in simple words,
``polyominoes with no holes''), see for example 
\cite{altrolibro}. The same questions as
above (determination of the $A_n$'s, asymptotics, critical exponents
for observables like the perimeter and the radius of gyration,\ldots)
apply to this subfamily, and involve in principle a different set of
critical exponents.


The point of this paper is that one can consider some measure of
interest $\mu_n(P)$ over polyominoes $P$ of size $n$, instead that the
uniform one. Of course, for such a measure to be interesting, it shall
relate to some relevant probabilistic process. Again, this connects to
the notion of universality of critical phenomena, where modifying the
measure in such a way would correspond to ``couple'' the first model
to a second one, and tune again the parameters such that the system
becomes critical (of which a signal would be the fact that the natural
``Boltzmann'' series, i.e., the grand-canonical partition function,
has an algebraic singularity at $z=1$). An example of such a
philosophy comes from random planar maps. On one side, there is an
overwhelming evidence that critical exponents associated to maps
(asymptotics in the enumeration, scaling of distances, etc.) are
universal, that is do not depend on the precise local structure of the
map (for example, are the same for random triangulations, or for
quadrangulations, or for all maps altogether).  Furthermore, if one
consider maps ``with matter'' (that is, coupled with a critical
Statistical Mechanics model, such as the Ising Model, the Potts Model,
the $O(n)$ Loop Model, etc.), the critical exponents change, again in
a universal way, that depends only on the type of matter introduced,
and not on the local structure of the map. In some rare cases, the
introduction of matter may even simplify the problem (for example, the
enumeration of maps is much simpler if they are equipped with a
spanning tree,
which is the limit $q \to 0$ of the $q$-colour
Potts Model).


In the case of polyominoes, a simple example in this direction is the
measure induced by critical site percolation (that is, the $q$-state Potts model
in the limit $q\to 1$), e.g.\ on the triangular
lattice (which is the simplest case, as, by simple symmetry arguments,
it is known that the critical parameter is
$q_c=\frac{1}{2}$). Interestingly, this measure is much simpler to
study than the original problem, and is quite explicit: calling $b(P)$
the number of faces not in $P$, and adjacent to $P$, we have
$\mu_{n}^{\rm perc}(P)=2^{-n-b(P)+1}/n$ for polyominoes $P$ of size
$n$.  Also, in this case, exact sampling in polynomial time can be
perfomed quite easily: one should just explore the percolation cluster
containing the origin, repeating the algorithm up to have the desired
size, and perform anticipated rejection on small clusters.
The peculiar factor $1/n$ has a trivial explanation in this case: when
the underlying lattice is face-transitive (as is the case for the
square, hexagonal and triangular lattices, for example), without loss
of generality we can consider polyominoes rooted at one face, as there
is a 1-to-$n$ correspondence between unrooted and rooted objects.  In
particular the corresponding enumeration series is just $n A_n^{\cL}$,
and all statistical averages remain the same.

The measure induced by percolation is a simple illustration of how
modifications of the uniform measure induced by a Statistical
Mechanics model on the whole plane, although apparently more
complicated, may be more accessible than the uniform
measure. Exploring one certain class of examples within this
framework, namely the ones induced by the Abelian Sandpile Model (ASM)
of Statistical Mechanics \cite{dhar1} (which is related to Uniform
Spanning Trees, that is, the $q$-state Potts model in the limit 
$q\to 0$), is the topic of this paper. Contrarily to the model of
percolation (and, more generally, of critical $q$-colour Potts Model),
this model induces measures on polyominoes supported on the
simply-connected ones, that is, our (grand canonical) measures $\mu(P)$
will be non-zero if and only if the polyomino $P$ has no holes.

Other natural measures on lattice animals, with a large literature,
that we do not mention at length in this paper are for example the
\emph{Diffusion-Limited Aggregation} model (DLA) or the 
\emph{Eden Model} \cite{DLAprl,DLApra,eden, aggregrev}. These models
are, yet again, simpler versions of the uniform measure over
polyominoes, but, contrarily to the point stressed here, the
simplification does not come from the fact that the measure is defined
in terms of a Statistical Mechanics model, but rather from the fact
that the configurations can be generated by iteratively adding the
unit elements one by one, with some growth rule.



\section{Avalanches in the Abelian Sandpile Model and polyominoes}

\noindent
The \emph{Abelian Sandpile Model} \cite{btw} is a lattice automaton in
the class of out-of-equlibrium models in Statistical
Mechanics. Pictorially, it is a model in which some ``sand'' arrives
in the system, according to some protocol, and then the local
instabilities are relaxed through some ``sand avalanches'', which are
possibly large, so that the sand can ultimately leave the system
through its boundary.  When a single grain of sand is added, provided
that an avalanche occurs, every site has performed either a positive
number of topplings, or none, and the set of sites which have
performed at least one toppling is connected, and thus constitutes a
non-empty polyomino.  Here we shall give a short introduction to the
formalism, following in part the notations of \cite{dhar1,noicipro}.

By the celebrated work of Dhar and collaborators
\cite{dhar1,dhar2,dhar3,dhar4}, it is known that, under the protocol in
which the sand is added randomly and uniformly, the steady-state
probability distribution of the sand configurations is supported on
the so-called ``recurrent configurations'', and is uniform. Also, the
uniform measure is stable under addition of any given configuration,
followed by relaxation.  These configurations are characterised by the
avoidance of an infinite list of ``forbidden subconfigurations''
(FSC), and are in bijection with the spanning trees of the lattice,
rooted at the boundary, through a (slightly non-canonical\footnote{The
  bijection is described in terms of an auxiliary data structure: for
  each site, one shall choose a total ordering of the incident
  edges.}) algorithm called ``burning test''. The relation between
configurations and spanning trees is valid if we consider the boundary
as a single site. If instead we prefer to keep a visual notation
induced by the lattice, and do not connect the boundary edges among
themselves, it is more precise to say that the relation is with rooted
spanning forests, where each component of the forest is rooted at a
boundary edge.  Yet another characterisation of recurrent
configurations is that, by adding a ``frame identity'' to the
configuration and performing the resulting avalanches, the system goes
back to the original configuration, and the avalanche consists in
exactly one toppling per site (the frame identity $\framid$ is the
configuration such that $\framid(v)$ is the number of boundary edges
incident on $v$).

It is useful to recall the main ideas of the Propp and Wilson LERW
algorithm \cite{proppwilson} for the exact sampling of rooted spanning
trees, or more generally rooted spanning forests. The algorithm, for a
generic graph with boundary edges, goes as follows. Choose any
ordering of the sites of the domain (excluding the boundary).
Initialise the \emph{absorbing set} to the boundary. Then, for every
site, if it is not already in the absorbing set, start a random walk
from the site (with rates associated to the Laplacian matrix of the
graph), up to reaching the absorbing set, and add to the absorbing set
the loop-erasure of this walk (performed in the time ordering of the
walk).  At the end of the algorithm we have a rooted spanning forest,
with roots on the initial absorbing set, uniformly sampled, and in
bijection with recurrent configurations through the burning test.

From the point of view of Statistical Mechanics, the most natural
measure on sand configurations is the uniform measure on recurrent
configurations.  From this point onward, our constructions will be
tacitly assumed to be performed over sand configurations sampled with
this measure.  

Some reflection shows that, for an avalanche to produce
a non-simply-connected polyomino, it shall surround a FSC, thus the
measure on polyminoes induced by avalanches on uniform random
recurrent configurations is supported on the simply-connected
subfamily. This remark is implicit in the work of Dhar, and appears
explicitly for example in \cite{redig}.

In general, avalanches may involve more than one toppling on certain
sites, a well-known fact which has led to the definition of ``waves of
avalanches'', in \cite{priez94}.  The characterisation of recurrent
configurations has an implication on the wave decomposition.  Indeed,
for any recurrent configuration $z$, the relaxation of $z+\framid$
gives again $z$, through an avalanche that makes each site topple
exactly once. As a result, for every portion of the frame identity,
$0 \prec u \prec \framid$, the relaxation of $z+u$ must produce an
avalanche that makes each site topple either one or zero times, and
the support of sites which have not toppled must remain accessible
from the boundary, as they will be toppling if we now add $\framid-u$
to the configuration and relax. In other words, if we add the amount
of sand described by $0 \prec u \prec \framid$, the resulting
avalanche will contain no more than a single wave.  We shall call
\emph{boundary avalanche} an avalanche induced by a $u$ of this form.

The study of the probability distribution of avalanches, and possibly
of the single waves, has been performed since the early days of the
model, but has proven difficult and controversial, and also
complicated to analyse on numerical experiments, because of strong
finite-size corrections
\cite{btwearly,manna,priez94,priez96,pacz,priez98}. Part of the
complicancy is due to the interplay among the different waves (cf.\ in
particular \cite{pacz}). It is thus conceivable that the study of
boundary avalanches does not suffer of the same pathologies as for
generic avalanches.

For definiteness, let us describe a process consisting of single-site
boundary avalanches, that we shall call the 
\emph{permutation boundary avalanche process}.
Let us call $V$ the number of sites in the domain (i.e.\ its
``volume''), $\cB \subset E$ the set of boundary edges, and
$B=|\cB|=|\framid|$ the number of boundary edges (which is also the
number of sand grains in the frame identity).  Let $\sigma \in
\mathfrak{S}_B$ be a random permutation of the boundary edges. We can
add the grains of sand constituting $\framid$ one by one, in the order
given by $\sigma$, and register the $B$ (possibly empty)
avalanches. By the abelianity properties of the ASM, the collection of
all the $B$ supports of the avalanches (i.e., the $B$ polyominoes)
coincides with the avalanche due to the addition of the whole frame
identity, and thus constitutes a partition of the domain. By the
stability of the uniform measure on recurrent configuration under
addition of deterministic configurations, for every $1 \leq k \leq B$,
the probability distribution over the polyomino associated to $b_k$,
the $k$-th boundary edge in the order of $\sigma$, is only a function
of the boundary edge itself, and not of the position it occupies in
the ordering $\sigma$.  In particular, if $v_{b}$ is the average size
of the polyomino associated to a boundary avalanche due to the
boundary edge $b$, we must have $\sum_b v_b = V$ (again, regardless of
the choice of $\sigma$).  In particular, on a lattice in which the
boundary edges are all equivalent (i.e., on ``boundary-edge-transitive
graphs''\footnote{That is, graphs $G$ with an outer boundary $\cB$
  s.t., for all $b_1,b_2 \in \cB$, there exists $g \in \aut(G)$
  s.t.\ $g(b_1)=b_2$.}), we must have $v_b=V/B$ for all $b$.

A strongly related process, that we shall call the 
\emph{BT boundary avalanche process}, is more directly related to the
burning test, and the Propp and Wilson algorithm for generating
uniform rooted spanning trees~\cite{proppwilson}.
In this case, for each site $v$ we shall choose, once and for all, a
total ordering $\cO$ of the set of incident edges. We shall now add
the whole $\framid$, and perform the relaxation in parallel. The sand
grains of $\framid$ are ``coloured'' in $B$ different colours. Each
site $v$ will become unstable at some moment of the avalanche, that
is, it will have a height $h+c$, where $k$ is the maximal allowed
stable height, and $1 \leq c \leq d(v)$. The colour of the site $v$ is
inherited from the colour of the site $u$ which has donated the $c$-th
grain of sand among those which have been donated to $v$ at the
present stage of the avalanche, where ``the $c$-th'' neighbour is
defined according to the given ordering $\cO$. We can visualise the
process of colour inheritance by drawing an oriented edge $(uv)$ in
this case. The overall set of oriented edges added in this way
describes the rooted spanning forest which, through the burning test,
is in bijection with the given recurrent configuration. And, as we
have mentioned, the partition of the domain into polyominoes can be
studied in terms of the components of the forest obtained through the
Propp and Wilson algorithm.

We must have some values $v'_{b}$ (in principle different from the
$v_{b}$'s) for the average size of the polyomino associated to the
tree rooted on the boundary edge $b$, with 
$\sum_b v'_b = V$.  The independence of the set of spanning forests
form the choice of ordering $\cO$
implies that the $v_{b}$'s do not depend on $\cO$, and thus, in particular,
on a boundary-edge-transitive graph, we have $v'_b=V/B$ for all $b$.

\begin{figure}[tb!]
\[
\begin{array}{lp{41pt}p{41pt}p{41pt}p{41pt}p{41pt}p{41pt}p{41pt}}
\multicolumn{8}{l}{\includegraphics[scale=.85]{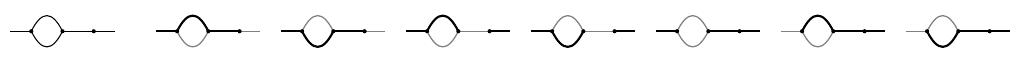}}\\
\rule{63pt}{0pt} & (3,0) & (3,0) & (2,1) & (2,1) & (1,2) & (0,3) & (0,3) 
\end{array}
\]
\[
\begin{array}{p{41pt}|p{41pt}p{41pt}p{41pt}p{41pt}p{41pt}p{41pt}p{41pt}}
$z$
   & \hspace{3pt}221 & \hspace{3pt}220 & \hspace{3pt}211 & \hspace{3pt}210 & \hspace{3pt}201 & \hspace{3pt}121 & \hspace{3pt}021 \\
\hline
$\sigma=[1,2]$ & (3,0) & (2,1) & (3,0) & (2,1) & (1,2) & (0,3) & (0,3) \\
$\sigma=[2,1]$ & (0,3) & (3,0) & (2,1) & (3,0) & (2,1) & (0,3) & (1,2) \\
\end{array}
\]
\caption{\label{fig.esempiocorrisp}An example of the correspondence
  between the BT boundary avalanche process and the permutation
  boundary avalanche processes for the possible choices of $\sigma$,
  for the small graph with $V=3$ and $B=2$ depicted on the top-left
  corner. Top: the list of the 7 spanning forests, and the
  corresponding list of $(|T_1|,|T_2|)$. Bottom: the 7 recurrent
  configurations, and the associated lists of $(|P_1|,|P_2|)$ for the
  2 permutations of the boundary edges. The three unordered lists are
  the same (namely, $(3,0)$, $(2,1)$, $(1,2)$ and $(0,3)$ are repeated
  2,2,1,2 times, respectively), this being the consequence, for this
  graph, of the statement that the permutation boundary avalanche
  process and the BT boundary avalanche process on the uniform measure
  over recurrent configurations are the same probabilistic process.}
\end{figure}

A crucial non-trivial fact is that the permutation boundary avalanche
process and the BT boundary avalanche process are in fact 
\emph{the same probabilistic process}.
A way of seeing this is to realise that in the permutation boundary
avalanche process, for any given $\sigma$, we can construct some trees
on the various avalanches, following the rules of the burning
test. Conditioning the sand configuration $z$ to have some avalanche
support $P=P_{b_1}$ for the boundary avalanche associated to the
boundary edge $b_1$ corresponds to say that $z|_P$ is recurrent for an
ASM model defined on a suitable restriction of the domain to
$G \smallsetminus P$, with appropriate boundary conditions, and that
the heights in the sites adjacents to $P$ are such that, after the
topplings on $P$ have been performed, no site has reached its critical
height value (this condition can be rephrased by a shift of both the
height values and the critical height values at these sites). We can
use this argument repeatedly, for all $b$ in $\cB$ in the order given by
$\sigma$, to deduce that the spanning forests constructed from the
permutation boundary avalanche process for the given $\sigma$, applied
to the list of all recurrent configurations, produce the list of all
spanning forests on the domain, with no repetitions. In particular,
$v'_b=v_b$ for all $b$, and more generally we can calculate any
observable for one process using the defining properties of the other
process (we will use this argument several times in the following
sections). See Figure \ref{fig.esempiocorrisp} for an illustration.

A typical example of boundary-edge-transitive domain is a 
$L_x \times L_y$ cylinder, in which (say) $L_x$ is the periodicity and
$L_y$ is the distance between the two portions of the boundary. In
this case $V/B=L_y/2$, and thus is a divergent quantity if we perform
the thermodynamic limit $V \to \infty$ by keeping the aspect ratio
fixed.  We shall call this case the \emph{cylinder geometry}.  A
variant of this geometry is again a $L_x \times L_y$ cylinder, but
now, instead of having two open boundaries, we have an open boundary
and a ``folded'' boundary, that is, a toppling at $(i,j)$ on this
boundary leaves one particle at $(i,j)$, and gives out three
particles, in the directions W,S,E.  We shall call this case the
\emph{folded cylinder geometry}. Note that the folded geometry can be
interpreted as an ordinary geometry $L_x \times 2L_y$, where we
restrict to configurations which are symmetric under horizontal
reflection (and add the sand to the system accordingly).

\begin{figure}[tb!]
\begin{center}
\rotatebox{180}{\includegraphics[scale=.25]{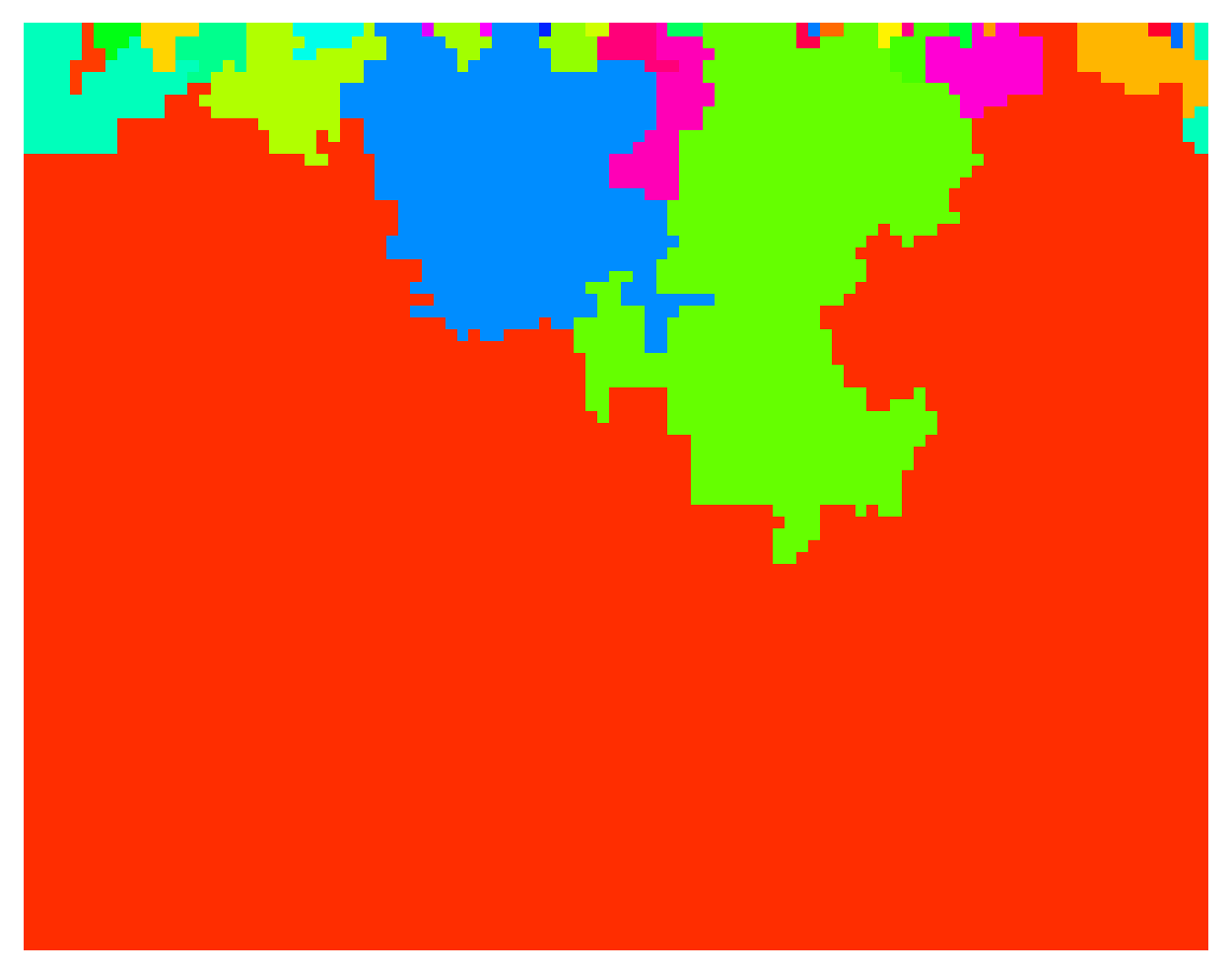}}\;%
\rotatebox{180}{\includegraphics[scale=.25]{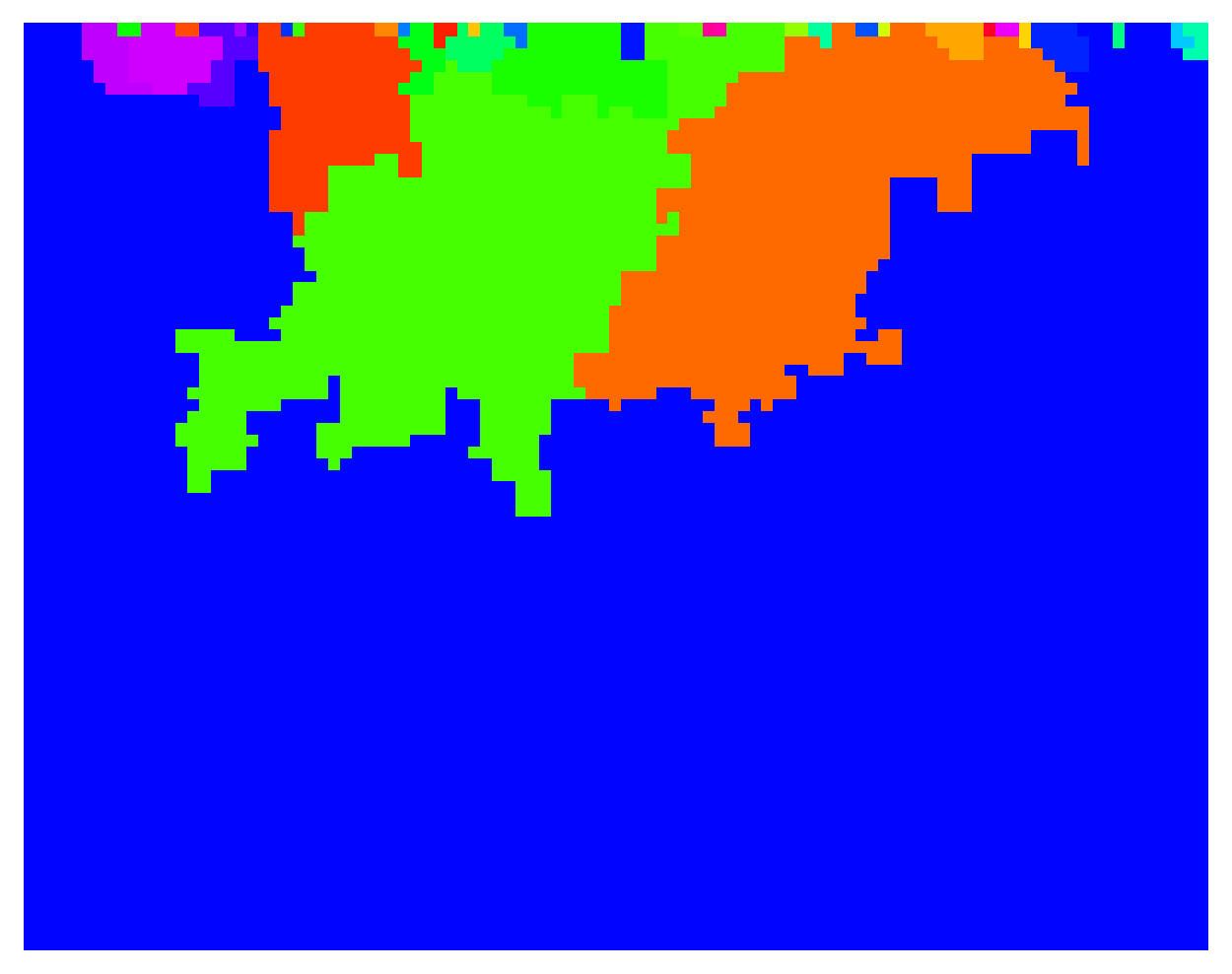}}\;%
\rotatebox{180}{\includegraphics[scale=.25]{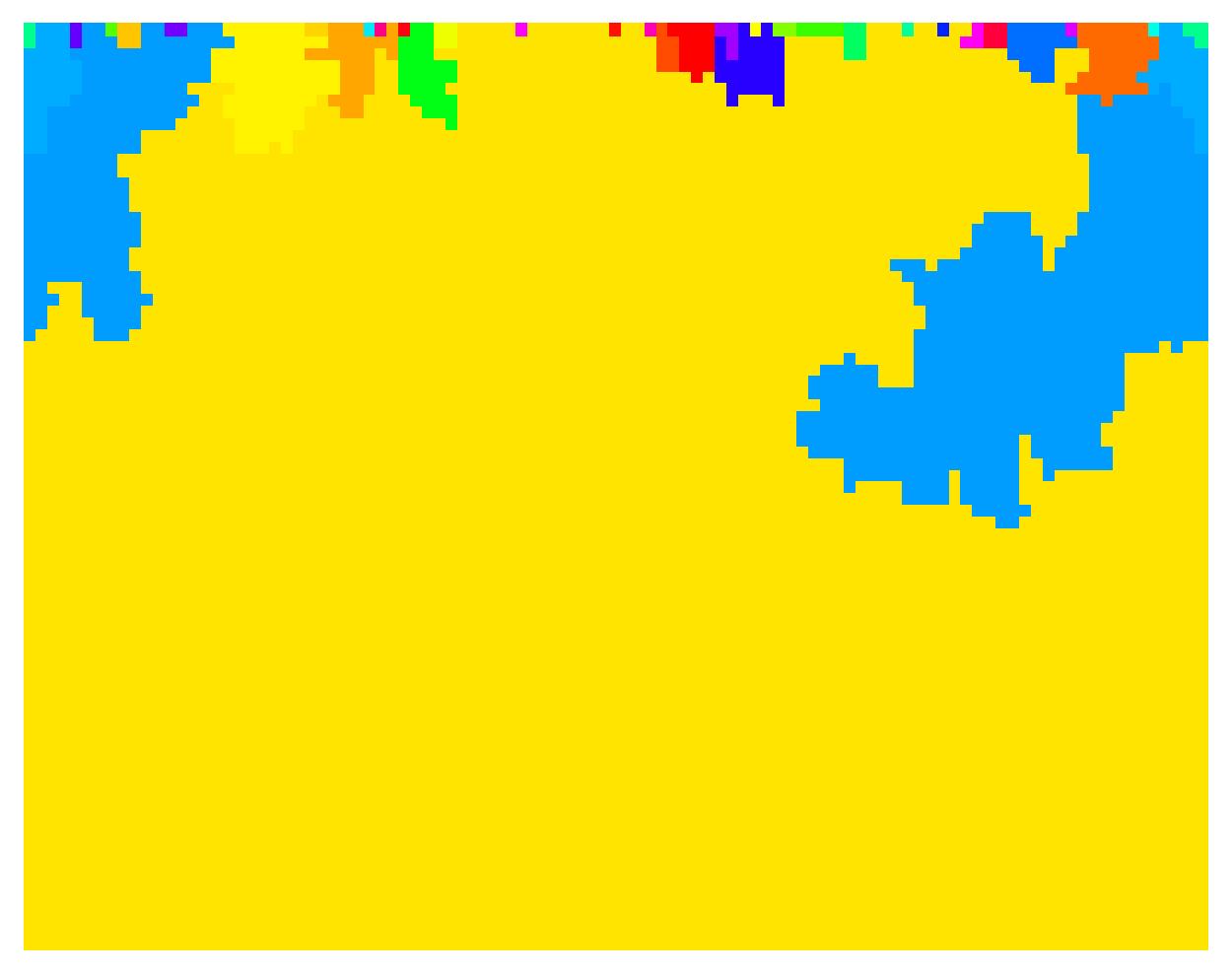}}
\\
\rotatebox{180}{\includegraphics[scale=.25]{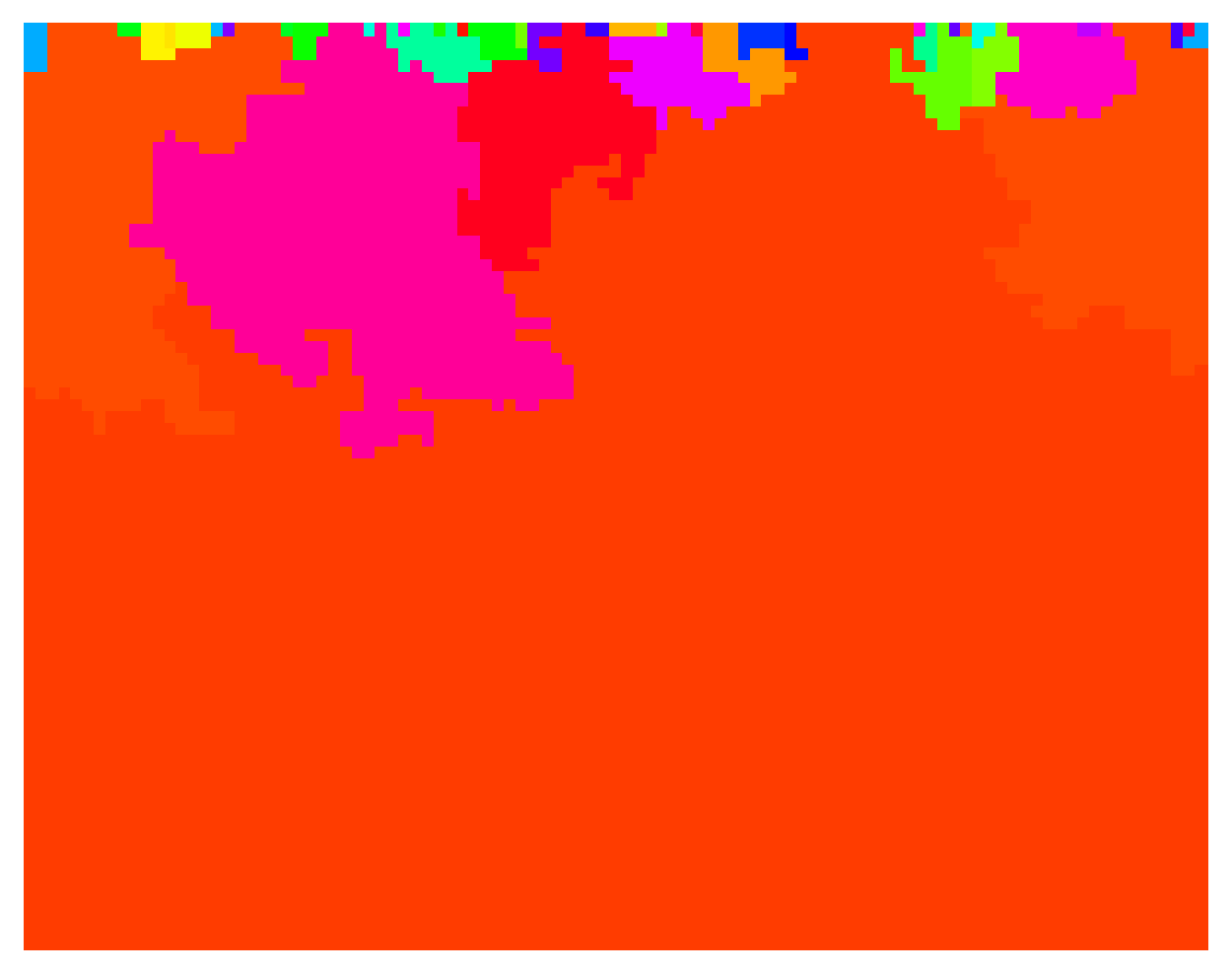}}\;%
\rotatebox{180}{\includegraphics[scale=.25]{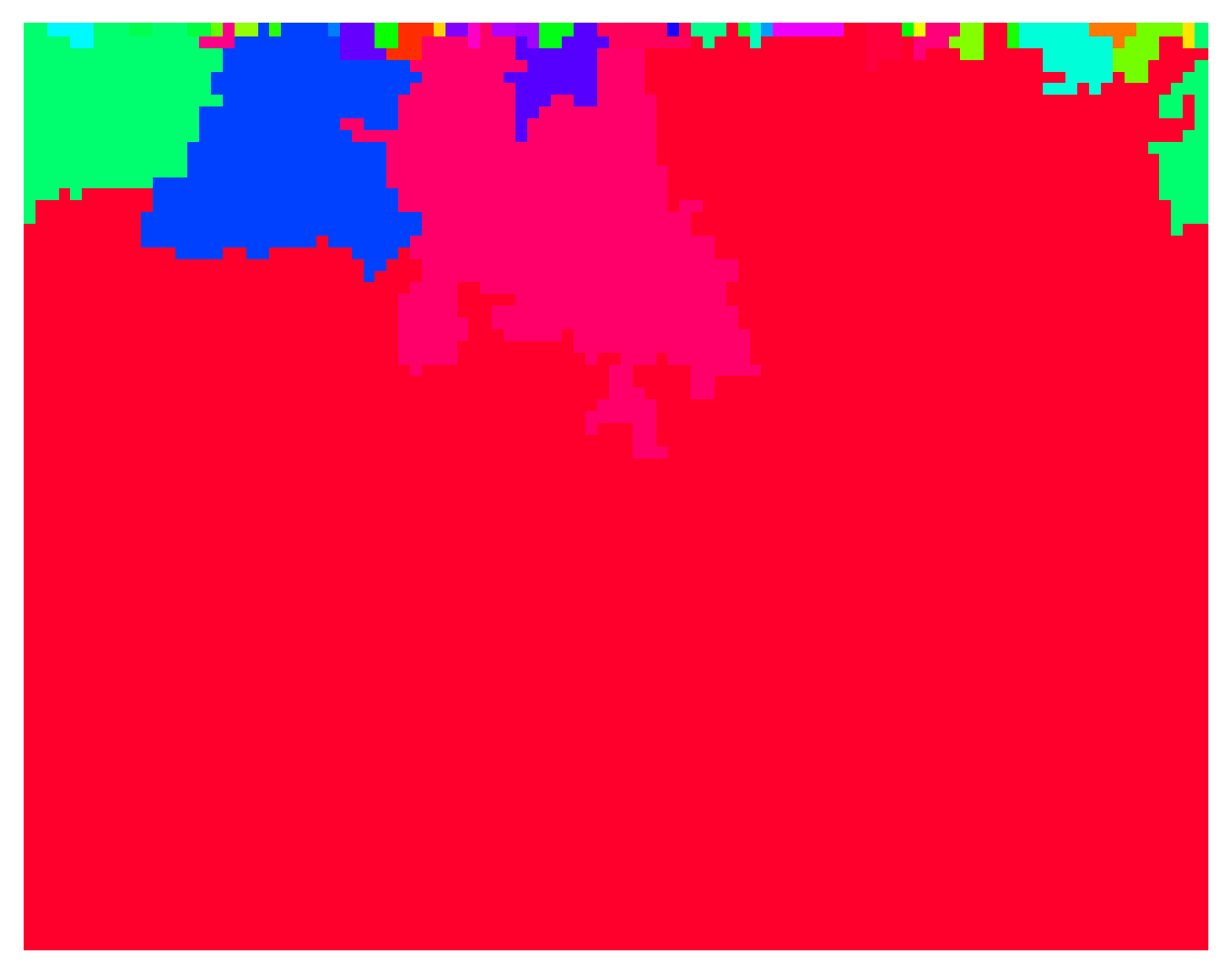}}\;%
\rotatebox{180}{\includegraphics[scale=.25]{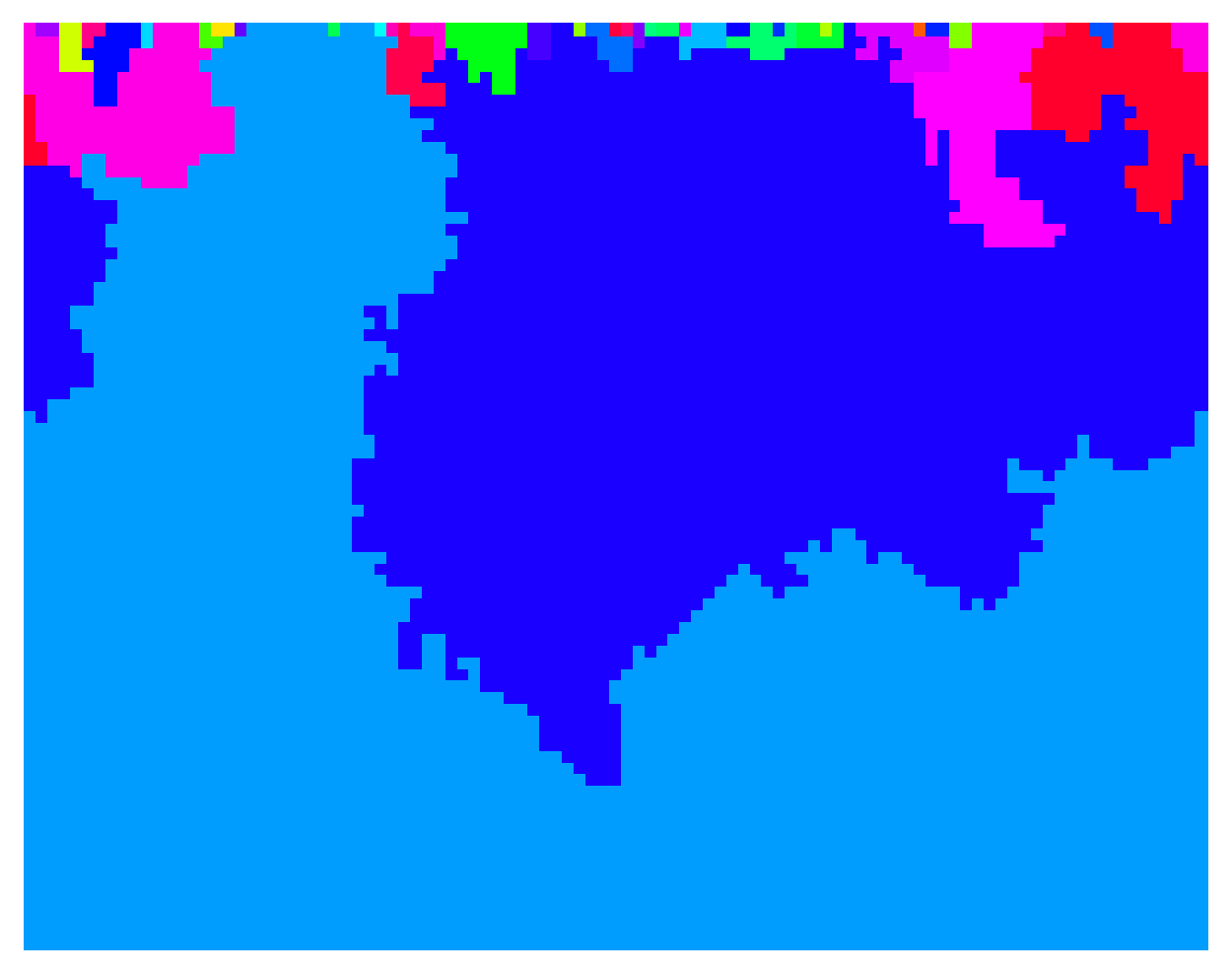}}
\end{center}
\caption{\label{fig.exproce} Some examples of realisations of the
  permutation boundary avalanche process on the square lattice in a
  folded-cylinder geometry with $L_x=101$ (only the most relevant part
  of the cylinder is shown). The hue value of the colour describes the
  position of the corresponding boundary edge in the permutation
  $\sigma$.}
\end{figure}

Some examples of realisations of this process are given in Figure
\ref{fig.exproce} at the end of this paper.



\section{Some accessible observables in the Boundary Avalanche Process}

\noindent
In this section we want to evaluate some statistical observables in
the Boundary Avalanche Process. The key idea is that we can use the
bijection between the implementation of the burning test and the
construction of spanning forests rooted at the boundary edges.  Then,
we can use either the implications of the Propp and Wilson LERW
algorithm \cite{proppwilson}, or also, more directly, the Kirchhoff
Matrix-Tree Theorem, by evaluating determinants of suitable Laplacian
matrices. Not surprisingly, these probabilities will turn out to be
ratios of determinants of very similar matrices, so that in fact, by
the Jacobi's theorem on complementary minors, through ``small''
determinants involving the inverse of the Laplacian matrix (that is,
the Green's function).

Note however that not all the potentially useful observables can be
calculated directly by this method. For reasons reminiscent of the
Lindstr\"om--Gessel--Viennot lemma, or the Kasteleyn solution of the
Dimer Model on bipartite planar graphs, probabilities of events are
accessible only if some topological property of the event guarantees
that the signs appearing in the determinant are controlled.

A useful formalism goes through Grassmann calculus, that is, a
representation of determinants (and determinants of minors) as formal
Gaussian integrals over complex scalar non-commuting variables, as
described in detail in \cite{itzdro1,noialan}. In this case, the roots
$R=\{r_i\}$ of the forests are described by factors 
$\psibar_{r_i} \psi_{r_i}$ in the integrand, while the factor
$\psibar_{u_1} \psi_{v_1} \cdots \psibar_{u_k} \psi_{v_k}$ implements
the fact that the vertices in some ordered list $U=(u_1,\ldots,u_k)$
are connected pairwise to the vertices in the list
$V=(v_1,\ldots,v_k)$ (according to some permutation 
$\sigma \in \mathfrak{S}_k$, that is, $u_j$ is in the same component
than $v_{\sigma(j)}$, and is not in the same component of any other
$u_i$, or $v_i$, or $r_i$). However, such an event comes with a sign
equal to the signature of $\sigma$. That is, for the three lists
$R=\{r_1,\ldots,r_h\}$, $U=(u_1,\ldots,u_k)$ and $V=(v_1,\ldots,v_k)$,
we will consider Grassmann integrals of the form 
\be 
Z_{R,U,V} = \int
\cD(\psi,\psibar) \; \Big( \prod_{r \in R}\psibar_r \psi_r \Big) \;
\psibar_{u_1} \psi_{v_1} \cdots \psibar_{u_k} \psi_{v_k} \; e^{\psibar
  L \psi} 
\ef.
\ee 
%
The consequence of the
Kirchhoff Theorem is that these expressions count (with signs) certain
$h+k$-component spanning forests of the graph,
\be 
\label{eq.2877646374}
Z_{R,U,V} = \sum_{\substack{
F=\{T_1,T_2,\ldots,T_{h+k}\} \subseteq G \\
r_i \in T_i \ \textrm{for}\ 1\leq i \leq k\\
u_i, v_{\sigma(i)} \in T_{k+i} \ \textrm{for}\ 1\leq i \leq h}}
\epsilon(\sigma)
\ef.
\ee
The explicit calculations on a generic weighted digraph $\cG$ (with a
boundary), such that the sum of the weights of the outgoing edges of a
vertex is the same for all vertices, involve the graph Green's
functions $G(u,v)$, identified by the defining equation
$L_u G(u,v) = \delta_{u,v}$, where 
$L_u f(u) = \sum_{(uu')} w_{(uu')} (f(u')-f(u))$ is the graph
(weighted) Laplacian (w.r.t.\ position $u$). The collection of the ``boundary
Green's functions'' $G(u,v)$, for $v$ on the boundary of $\cG$,
corresponds to the probabilities that a random walk, starting at $u$,
diffusing with the weights $w_e$ and absorbed at the boundary,
terminates in $v$. Thus, in particular, 
\be
\label{eq.2764367}
\sum_{v\in \partial \cG} G(u,v)=1
\qquad \forall \ u
\ef.
\ee
These remarks are of interest here because, as we will see, most of
the interesting choices of $(R,U,V)$ in (\ref{eq.2877646374}) are such
that (say) $U \cup R = \partial \cG$, so that the relevant Green's
functions in the evaluation of $Z_{R,U,V}$ are indeed boundary Green's
functions in the sense above.

The calculations are more explicit on portions of regular lattices,
and involve lattice sums on certain lattice Green functions on the
domain, which, when the domain allows for the use of the ``method of
images'', can be constructed in terms of the lattice Green function of
the infinite lattice under investigation (most notably, the square,
triangular or hexagonal lattice). The theoretical investigation of
lattice Green function has a long history, of which a breakthrough
result is due to L\"uscher and Weisz \cite{luwei} (where an important
ingredient is an observation of Vohwinkel unpublished elsewhere),
which, for the square and triangular lattice, has been implemented in
\cite{dongss1,dongss2} and in \cite{claudia}, respectively (recall
that, as polyominoes are defined on the faces of the lattice, the
Green function of the triangular lattice in fact relates to
polyhexes). See also \cite{capo084} for further details.

In order to calculate the algebraic asymptotic decay of probabilities
of events, however, it is enough to use the asymptotic Green function,
which for all lattices, once that the lattice spacing is rescaled in
order to have unit density, is universally
$G(\vec{x}_1,\vec{x}_2)=\frac{1}{4\pi} \ln |\vec{x}_1-\vec{x}_2|^2$.
However, in the special case of a straigth boundary, the method of
images implies that (say, for the square lattice) we have to consider
the combination
$G(\vec{x}_1,\vec{x}_2)-G(\vec{x}_1,\vec{x}_2-2\hat{e}_y)$, that
scales as $G_{\rm bd}(x,y)=\frac{1}{\pi} \frac{y}{x^2+y^2}$ for
$x^2+y^2 \gg 1$ (for the triangular lattice with unit density, we have
a correction factor 
$\aa=2^{\frac{1}{2}}3^{-\frac{1}{4}}$).


A first warm-up example of observable can be the explicit check of the
simple fact that any site must be in some tree of the forest.  So
we must have the identity
\be
Z_{\cB,\varnothing,\varnothing}=
\sum_{b \in \cB} Z_{\cB\setminus b,(b),(s)}
\qquad
\forall\ s
\ef.
\ee
On a generic graph $\cG$, and using the Kirchhoff Matrix-Tree Theorem
and Jacobi minor formula, this is rephrased into the statement
(\ref{eq.2764367}) above (and indeed the random walk defining the
boundary Green's function can be interpreted as the support for the
first LERW in Propp and Wilson's algorithm, when $s$ is chosen to be
the first vertex in the ordering).

It is instructive to check that, for the specific case of the square lattice and
in a limit of $y \gg 1$, $L_x, L_y \gg y$,\footnote{In this limit we can use the asymptotic form of the boundary
Green's function given above, and, as it will be useful only later on, trade lattice derivatives with ordinary
derivatives.} the
combinatorial statement above is in agreement with the identity
\be
\sum_{x \in \bZ}
\frac{1}{\pi} \frac{y}{x^2+y^2}
\simeq
\int_{-\infty}^{\infty}
\dx x \;
\frac{1}{\pi} \frac{y}{x^2+y^2}
=
1
\qquad 
\forall\ y
\ef,
\ee
(for the triangular lattice, a factor $\aa^{-1}$ for the density of
sites along a row cancels out with
the scaling factor $\aa$ in the Green's function).


A more interesting calculation consists (for example, in the case of
hexagonal cells) in determining the probability that the vertex in
$(x,y)$ is a triple point of the process, that is, its three adjacent
hexagonal faces are in three different polyominoes. The fact that 3 is
an odd number, that the set $U$ is on the outer boundary and the set
$V$ consists of adjacent faces implies that the annoying signs are in
fact protected, that is, of the six possible permutations, only the
three connectivity patterns with equal signature are allowed.  For
$y \gg 1$ the probability that $(x,y)$ is a triple point, and the
three adjacent polyominoes are rooted on the points $(x_1,x_2,x_3)$,
is given (up to a simple scaling factor for the lattice spacings, and
in a limit $L_x, L_y \gg x,y,x_1,x_2,x_3$) by the determinant of the
matrix
\be
M[x,y,(x_1,x_2,x_3)]=\left(
\frac{y}{(x_i-x)^2+y^2}\ ,\ \deenne{x}{{}}\frac{y}{(x_i-x)^2+y^2}\ ,\  \deenne{y}{{}}\frac{y}{(x_i-x)^2+y^2} 
\right)_{i=1,2,3}
\ee
Integrating over $x$ and $y$ gives the overall probability that 
the polyominoes rooted on the real axis at coordinates $(x_1,x_2,x_3)$ share a triple
point. A calculation shows that this probability is proportional to
the inverse of the Vandermonde factor,
\be
\int_{-\infty}^{\infty}
\dx x \;
\int_{0}^{\infty}
\dx y \;
\det
M[x,y,(x_1,x_2,x_3)]
\propto
\frac{1}{(x_3-x_2)(x_3-x_1)(x_2-x_1)}
\ef.
\ee
Integrating over the $x_i$'s, at $x=0$,
gives the overall probability that 
$(0,y)$ is a triple point, which is, for $y$ large enough,
\be
\int_{-\infty \leq x_1 \leq x_2 \leq x_3 \leq \infty}
\dx x_1 \dx x_2 \dx x_3
\det
M[0,y,(x_1,x_2,x_3)]
=
\frac{1}{2\pi y^2}
\ef.
\ee
Indeed, the algebraic decay $1/y^2$ is integrable at infinity, a fact
in agreement with the deterministic information that there are exactly
$L_x-2$ triple points in a configuration on a folded cylinder, that
is, asymptotically on average one triple point per column.

Now we calculate an observable in which the role of the signs is more
subtle. Consider a realisation of the boundary avalanche process, in a
limit $L_x, L_y \to \infty$, so that the boundary vertices can be
totally ordered along $\bZ$. For $i<j$, if the polyominoes $P_i$ and
$P_j$ share a boundary, then they have exactly two triple points, with
some polyominoes $P_{k_{\rm int}(i,j)}$ and $P_{k_{\rm ext}(i,j)}$. A
peculiar fact is that, of these two vertices, only one will be in the
range $\{i+1,\ldots,j-1\}$ (we will set it to be
$k_{\rm int}(i,j)$). So we can define unambiguously the vector 
$\vec{v}_{ij}=t_{i,j,k_{\rm ext}(i,j)}-t_{i,j,k_{\rm int}(i,j)}$,
where $t_{i,j,l}$ is the triple point between the polyominoes $P_i$,
$P_j$ and $P_l$, and set $\vec{v}_{ij}=0$ if $P_i$ and $P_j$ do not
share a boundary. Now, given two adjacent faces $v_1$, $v_2$, consider
$Z_{\cB \smallsetminus \{u_i,u_j\},\{u_i,u_j\},\{v_1,v_2\}}$. This
quantity gives the probability that $v_1 \in P_i$ and $v_2 \in P_j$,
minus the probability that $v_1 \in P_j$ and $v_2 \in P_i$. Call
$e'_{v_1,v_2}=(v'_1,v'_2)$ the oriented dual edge associated to the
oriented edge $(v_1,v_2)$. Remark that
\be
\sum_{(v_1,v_2)}
e'_{v_1,v_2}\;
Z_{\cB \smallsetminus \{u_i,u_j\},\{u_i,u_j\},\{v_1,v_2\}}
=
\eee\, \vec{v}_{ij}
\ef.
\ee
Indeed, the boundary between $P_i$ and $P_j$ is a polygonal curve resulting
from the concatenation of dual edges (in either orientation), going
from $t_{i,j,k_{\rm int}(i,j)}$ to $t_{i,j,k_{\rm ext}(i,j)}$. 

Similar arguments give, for a region $\Omega$,
\be
\sum_{(v_1,v_2) : v'_1 \in \Omega, v'_2 \not\in \Omega}
Z_{\cB \smallsetminus \{u_i,u_j\},\{u_i,u_j\},\{v_1,v_2\}}
=
\ppp\, (t_{i,j,k_{\rm int}(i,j)} \in \Omega, t_{i,j,k_{\rm ext}(i,j)}
    \not\in \Omega)
-
\ppp\, (t_{i,j,k_{\rm ext}(i,j)} \in \Omega, t_{i,j,k_{\rm int}(i,j)}
    \not\in \Omega)
\ef.
\ee
Calculations of the asymptotic behaviour of
observables of this form rely  on the
evaluation of the quantities 
$Z_{\cB \smallsetminus \{u_i,u_j\},\{u_i,u_j\},\{v_1,v_2\}}$, which
are related to the evaluation of the determinant of a matrix
of the form
\be
M'[x,y,(x_1,x_2)]=\left(
\frac{y}{(x_i-x)^2+y^2}\ ,\ \deenne{x}{{}}\frac{y}{(x_i-x)^2+y^2}
\right)_{i=1,2}
\ef.
\ee
In particular, taking as $\Omega$ the half-plane above height $y$, and
summing over all pairs $i<j$, we have
\begin{multline}
\label{eq.3876478}
\frac{1}{L_x}
\sum_{i<j}
\big(
\ppp\, (t_{i,j,k_{\rm int}(i,j)} \in \Omega, t_{i,j,k_{\rm ext}(i,j)}
    \not\in \Omega)
-
\ppp\, (t_{i,j,k_{\rm ext}(i,j)} \in \Omega, t_{i,j,k_{\rm int}(i,j)}
    \not\in \Omega)
\big)
\\
=
\int_{-\infty \leq x_1 \leq x_2 \leq \infty}
\dx x_1 \dx x_2 \dx x_3
\det
M'[0,y,(x_1,x_2)]
=
\frac{1}{\pi y}
\ef.
\end{multline}
Now the algebraic decay $1/y$ is not integrable at infinity, and gives
a sensible information on the fractal properties of the process. We
discuss the implications of this calculation in the next section.

\section{A scaling argument}

\noindent
We shall try to give a prediction for the asymptotic behaviour of the
tail of the probability distribution for the boundary avalanches.
Say that we are in a cylinder with aspect ratio of order 1.
Let us suppose that, on some length scales much larger than the
lattice spacing, and much smaller than the size of the domain, the
process of boundary avalanches is approximatively scale invariant.
Then, the distribution of the sizes of the avalanches must be a power
law for the range $1 \ll n \ll L^2$, and then must be truncated by the
finiteness of the domain, i.e.
\be
p_L(n) \sim \left\{
\begin{array}{ll}
n^{-\gamma} & n \ll L^2 \\
0           & n \gg L^2
\end{array}
\right.
\ee
The value of $\gamma$, unknown up to this point, can now be
determined: indeed we know
that $\sum_n n p_L(n) \sim L$, and such a behaviour is compatible with a
single value of $\gamma$, namely $\gamma=\frac{3}{2}$. 
In other words, we expect that $\ppp_L(n(P)>n) \sim n^{-\frac{1}{2}}$
for $1 \ll n \ll L^2$.

This sketchy prediction seems numerically verified, but somewhat ``for
the wrong reasons''. A more detailed description of the truly
scale-invariant process of boundary avalanches should be given in a
regime in which the geometry of the cylinder does not introduce a new
finite parameter in the model, that is, in a regime 
$L_y \gg L \gg 1$ (in this section it is convenient to adopt the
notation $L_x=L$).  In this case we do not see anymore the effect of
the top boundary of the cylinder, or a difference between the cylinder
and the folded-cylinder geometries, and we shall expect that there
exists almost surely one ``giant avalanche'', occupying a fraction
$1-\cO(L/L_y)$ of the volume, so that the probability distribution may
take the form (calling $V=L L_y$ the volume)
\be
\label{eq.distroansatz1}
p_L(n) \sim \left\{
\begin{array}{ll}
n^{-\gamma} & n \ll L^{2} \\
o(n^{-\gamma}) & L^{2} \ll n \ll V \\
p'_L(n)     & n = V-\cO(L^{2}) \\
0           & n > V
\end{array}
\right.
\ee
where $\sum_n p'_L(n) = 1/L$, and again we must have $\gamma>1$.
This implies for the average
\be
\frac{V}{L} = \eee(|P|)
= \frac{V-\Theta(L^{2})}{L} + \int^{L^{2}} \dx x \; x^{1-\gamma}
= \frac{V}{L}-\Theta(L)+\Theta((L^{2})^{2-\gamma})
\ee
which requires
\be
1=2(2-\gamma)
\ee
that is, again $\gamma=3/2$.

\begin{figure}[tb!]
\[
\makebox[0pt][l]{\rule{270pt}{0pt}$k$}
\includegraphics[scale=.75]{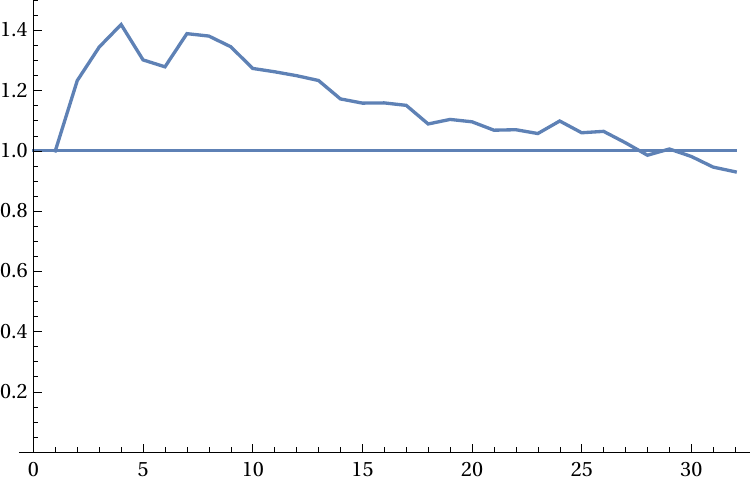}
\]
\caption{\label{fig.medie}Averages of the $k$-th largest polyomino in
  a process (except the giant one), multiplied by $k^2$, and rescaled
  so that the first value is 1, on the data presented in Figure
  \ref{distriProce}. A far-fetching conjecture based on the formula
  (\ref{eq.3876478}) would suggest that this function is 1, up to
  values of $k \ll L_x$.}
\end{figure}

It is not completely evident that, except for the trivial giant
avalanche, the process of boundary avalanches occupies a height of
order $L$ of the domain, and that the second largest avalanche is on a
scale $\sim L^2$. However, this can be established through the
calculation, performed in (\ref{eq.3876478}), of the average number of
interfaces between pairs of polyominoes that reach height $y$, which
scales as
$L/(\pi y)$.  So, this average goes from $\gg 1$ to $\ll 1$ when $y$
goes from much smaller than $L/\pi$ to much larger than $L/\pi$. As
avalanches have possibly fractal boundaries, but their interior has
Hausdorff dimension 2, we deduce that the second largest avalanche
must have a volume on the scale $\sim L^2$. Then, as the appearence of
each further avalanche approximately adds one to the number of
interfaces, from the behaviour in $L/y$ of our observable we may
deduce that the average sizes of avalanches listed in decreasing order
(and excluding the giant one) may form a sequence not too far from the
series $C L^2/k^2$, for some constant $C$, up to values of $k$ so that
the avalanches have macroscopic sizes. It is remarkable that such a
far-fetching prediction is vaguely in accordance with the numerics,
even at relatively small sizes (cf.\ Figure \ref{fig.medie}).

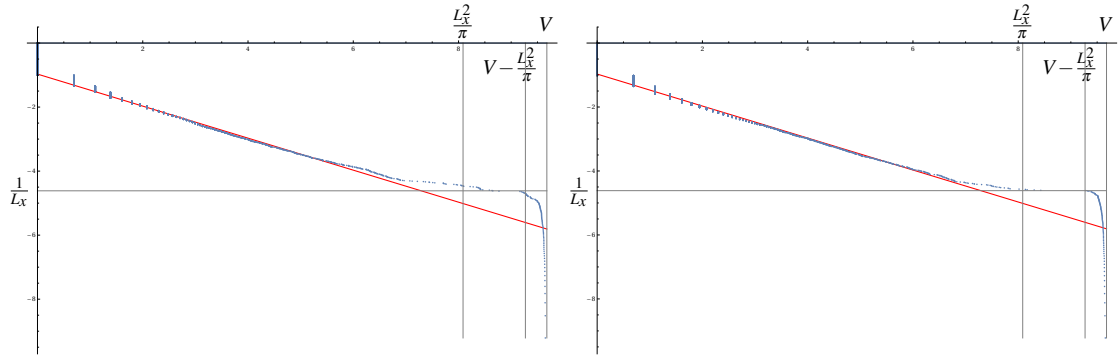
\begin{figure}[tb!]
\[
\makebox[0pt][l]{\setlength{\unitlength}{25pt}
\begin{picture}(1,0.001)
\put(-.4,2.4){$\scriptstyle{\frac{1}{L_x}}$}
\put(6.3,4.9){$\scriptstyle{\frac{L_x^2}{\pi}}$}
\put(6.8,4.25){$\scriptstyle{V-\frac{L_x^2}{\pi}}$}
\put(7.65,4.9){$\scriptstyle{V}$}
\end{picture}}
\includegraphics[scale=.25]{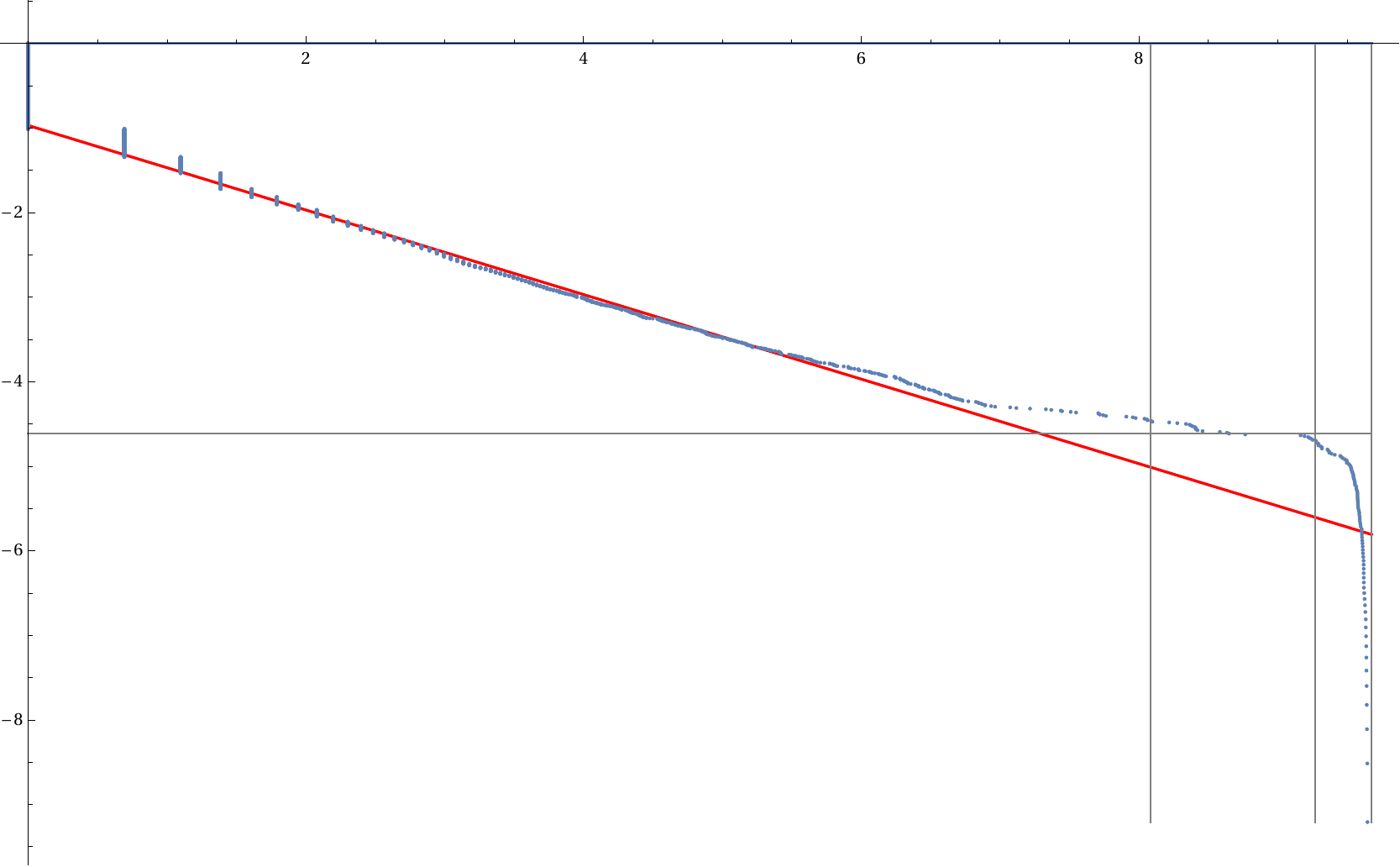}
\quad
\makebox[0pt][l]{\setlength{\unitlength}{25pt}
\begin{picture}(1,0.001)
\put(-.4,2.4){$\scriptstyle{\frac{1}{L_x}}$}
\put(6.3,4.9){$\scriptstyle{\frac{L_x^2}{\pi}}$}
\put(6.8,4.25){$\scriptstyle{V-\frac{L_x^2}{\pi}}$}
\put(7.65,4.9){$\scriptstyle{V}$}
\end{picture}}
\includegraphics[scale=.25]{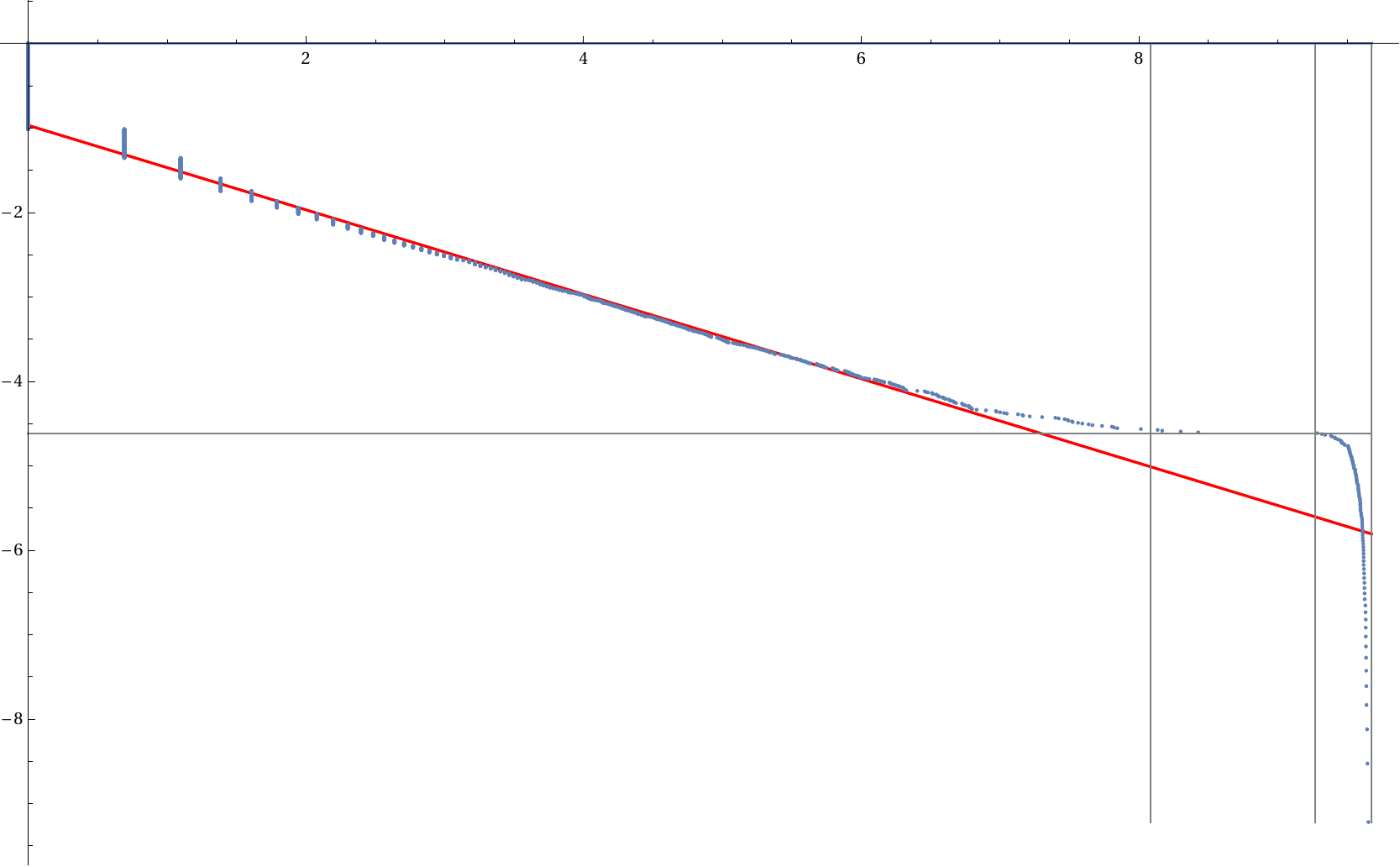}
\]
\caption{\label{distriProce}
Left: plot of the ordered list of $10^4$ avalanche sizes, for a
folded-cylinder geometry on the square lattice, of size $101 \times
158$.
We adopt a log-log plot, with a superposed red line of slope $-1/2$,
which highlights the validity of the ansatz in equation
(\ref{eq.distroansatz1}) in this case.
Right: plot of the ordered list of $L_x \times 10^2=10100$ avalanche
sizes, for $100$ realisations of the permutation process.}
\end{figure}

Note that, as yet another consequence of the properties of the ASM,
the probability distribution for the avalanche process, shown in the
bottom of Figure \ref{distriProce}, and for the single-site boundary
avalanches, shown in the top of the same figure,
are essentially coincident (except for the fact that the fraction of
giant avalanches in the first case is exactly equal to $1/L$, while in
the second case it is only approximatively equal to this value, with
Gaussian fluctuations on a scale compatible with an approximation of
independent events).  Indeed, as explained above, the coincidence of
these two distributions is implied by the principles of the Abelian
Sandpile Model, while the scaling ansatz only concerns the
determination of the qualitative properties of this function.

Arguments of this type have been a \emph{leitmotif} of this paper: the
relation between apparently different boundary avalanche processes has
allowed us to deduce fine statistical properties for each of them (and
in particular for the most basic procedure, of a single-site boundary
avalanche), by using each time the most convenient
formulation. Without using this multiplicity of definitions, we
wouldn't have been able to perform most of our calculations.


\nocite{*}
\bibliographystyle{eptcs}
\bibliography{generic}
\end{document}